\documentstyle[sprocl,textcomp,epsfig]{article}

\def\bea{\begin{eqnarray}}
\def\eea{\end{eqnarray}}
\newcommand{\MeVcc}   {\mbox{$ {\mathrm{MeV}}/c^2                          $}}
\newcommand{\GeVcc}   {\mbox{$ {\mathrm{GeV}}/c^2                          $}}
\newcommand{\eeto}    {\mbox{$ {\, \mathrm e}^+ {\mathrm e}^- \to             $}}
\def\MXN#1{\mbox{$ M_{\tilde{\chi}^0_#1}                                $}}
\def\XN#1{\mbox{$ \tilde{\chi}^0_#1                                     $}}
\newcommand{\smu}     {\mbox{$ \tilde{\mu}                                 $}}
\newcommand{\sel}     {\mbox{$ \tilde{\mathrm e}                           $}}
\newcommand{\sle}     {\mbox{$ \tilde{\ell}                                $}}
\newcommand{\stau}    {\mbox{$ \tilde{\tau}                                $}}
\newcommand{\GeV}     {\mbox{$ {\mathrm{GeV}}                              $}}
\def    \misspt      {\ifmmode{/\mkern-11mu p_t}\else{${/\mkern-11mu p_t}$}\fi}
\newcommand{\msmu}    {\mbox{$ M_{\tilde{\mu}}                             $}}
\newcommand{\mstau}   {\mbox{$ M_{\tilde{\tau}}                            $}}
\newcommand{\msle}    {\mbox{$ M_{\tilde{\ell}}                            $}}
\newcommand{\sell}    {\mbox{$ \tilde{\mathrm e}_{\mathrm L}               $}}
\newcommand{\selr}    {\mbox{$ \tilde{\mathrm e}_{\mathrm R}               $}}
\newcommand{\WW}      {\mbox{$ {\, \mathrm W}^+{\mathrm W}^-                  $}}
\newcommand{\GeVc}    {\mbox{$ {\mathrm{GeV}}/c                            $}}
\newcommand{\Ecms}    {\mbox{$ E_{\mathrm{\small cms}}                      $}}
\newcommand{\Lum}{${\cal L}\;$}
\newcommand{\msel}    {\mbox{$ M_{\tilde{\mathrm e}}                       $}}
\newcommand{\Mvis}    {\mbox{$ M_{\mathrm{\small vis}}                      $}}
\newcommand{\dgree}   {\mbox{$ ^\circ                                      $}}

\begin{document}

\title{RECONSTRUCTING SLEPTONS
IN CASCADE-DECAYS AT THE LINEAR COLLIDER}

\author{M. BERGGREN}

\address{LPNHE, Universit\'{e}s Paris 6 et 7, IN2P3--CNRS, Paris, France}
\maketitle

\abstracts{A method to reconstruct sleptons in cascade-decays at
the FLC is presented. It is shown that experimental mass-resolutions
as low as 8.7 $\MeVcc$ are attainable.}

{\it Work presented at the International Conference on Linear Colliders (LCWS04), 19-23 April 2004, Le Carre des Sciences, Paris, France.}
\vspace{1cm}

I have studied the cascade process $\eeto\XN{2}\XN{2}\to\sle\sle ll\to\XN{1}\XN{1}lll'l'$
at a number of SPS points[1,2]. Simulated
%was studied at a number of SPS points\cite{jag}\cite{sps}. Simulated
%events were processed by the fast detector simulation SGV\cite{susygen}\cite{tdr}\cite{sgv}.
events were processed by the fast detector simulation SGV[3-5].
Events that did not contain other charged particles than the four
leptons, and had clear signs of undetected particles were further
analysed. The $\tilde{{\chi}}^{0}$:s were reconstructed using the
seen leptons and conservation laws. $E$ and $|p|$ of the \XN{2}:s
follow from energy conservation and their direction is found the way
$p_{W}$ in $\eeto\WW\rightarrow l\nu l\nu$ is found, yielding two
%possible solutions\cite{wwtgc}
possible solutions[6].
The momentum of the \XN{1}
was obtained from $P_{\ell\ell}$ and each of the two $\XN{2}$ solutions,
and $P_{\tilde{\ell}}$ by adding the momentum of $\XN{1}$ to the
appropriate $P_{\ell}$. As there is no way to determine which lepton
came from the slepton decay, and there are two possible values for
the momentum of $\XN{1}$, there will be four possible solutions for
each slepton (or eight in 4$e$ or 4$\mu$ events). 
%\footnote{Similar ideas have been
%applied by other workers for background rejection in SUSY searches
%at LEP\cite{wwsusy}, to cascades at LHC\cite{nojiri} and to $\sel$-production
%in $e^{+}e^{+}$-collisions\cite{A-S}.} 
\footnote{Similar ideas have been
applied by other workers for background rejection in SUSY searches
at LEP[7], to cascades at LHC[8] and to $\sel$-production
in $e^{+}e^{+}$-collisions[9].} 

{\bf Slepton mass determination in SPS1a.} In SPS1a,
at \Ecms= 500 \GeV, and \Lum =500 fb$^{-1}$, 800 events of this
type are expected. There are no other cascades open, but one expects
10$^{6}$ other SUSY events. Events were selected as above: There
should nothing but four charged leptons seen, $\misspt>10\GeVc$,
\Mvis $\in${[}100, 300{]} \GeVcc, $E_{seen}$($E_{seen,neut}$)
$<$ 300 (150) \GeV. The thrust-axis should be above 0.3 Rad, and
$E_{calo,below 30\dgree}$$<150$ \GeV\ (the last two cuts removes the
$\gamma\gamma$ background). For all eight (or 16) possible values
of $P_{\tilde{\ell}}$ the invariant mass was calculated. A narrow
peak at the right mass corresponds to the correct solutions (two per
event). The decay of the \XN{2} is a three-body decay ($\ell\ell$
\XN{1}), so the tail from the wrong solutions can be suppressed by
demanding that they should be in the Dalitz-triangle, and be within
the narrow bands corresponding to the right combination (fig 1a).%
\begin{figure}[h]
\centering\includegraphics[%
  scale=0.18]{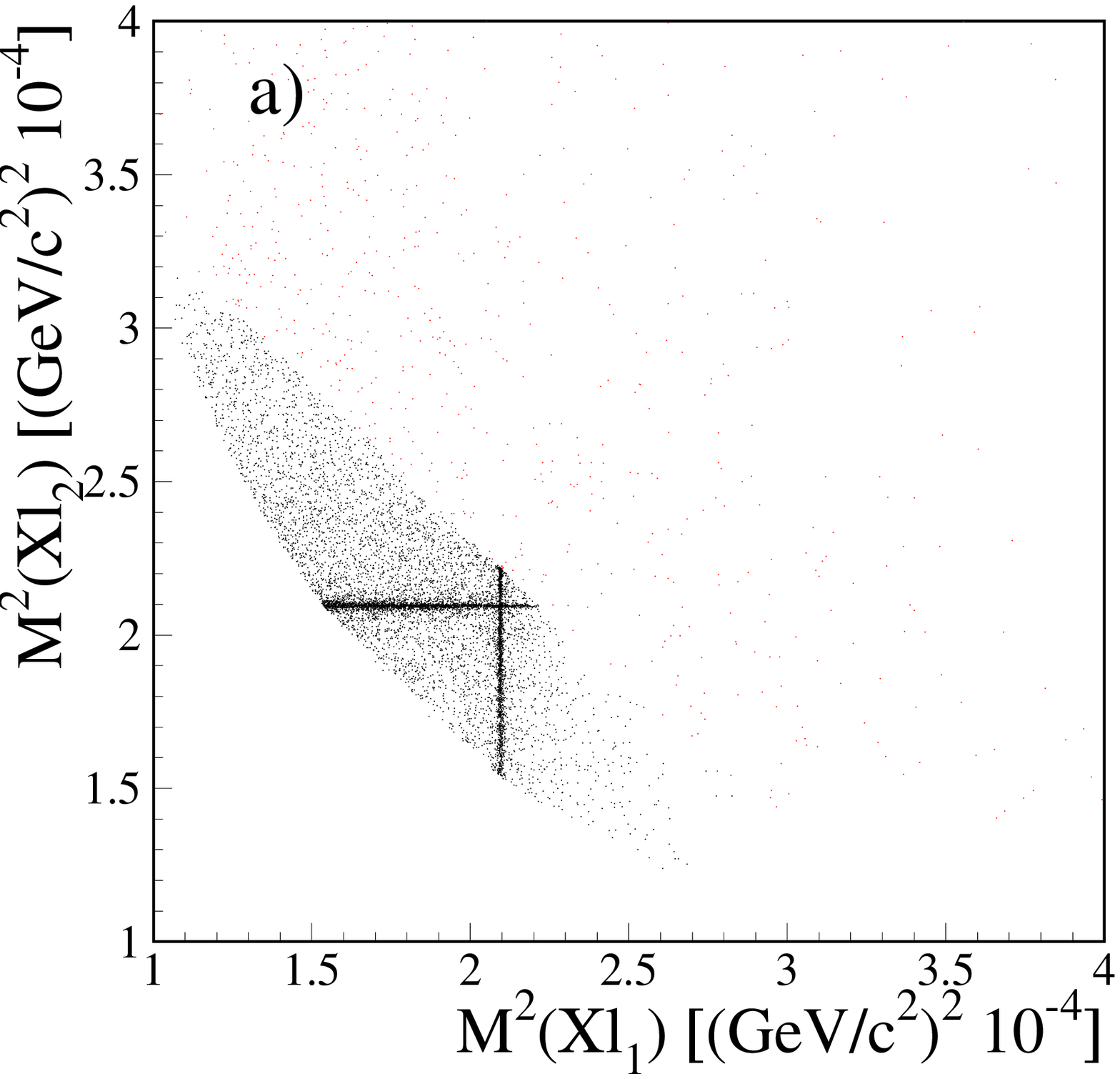}\includegraphics[%
  scale=0.18]{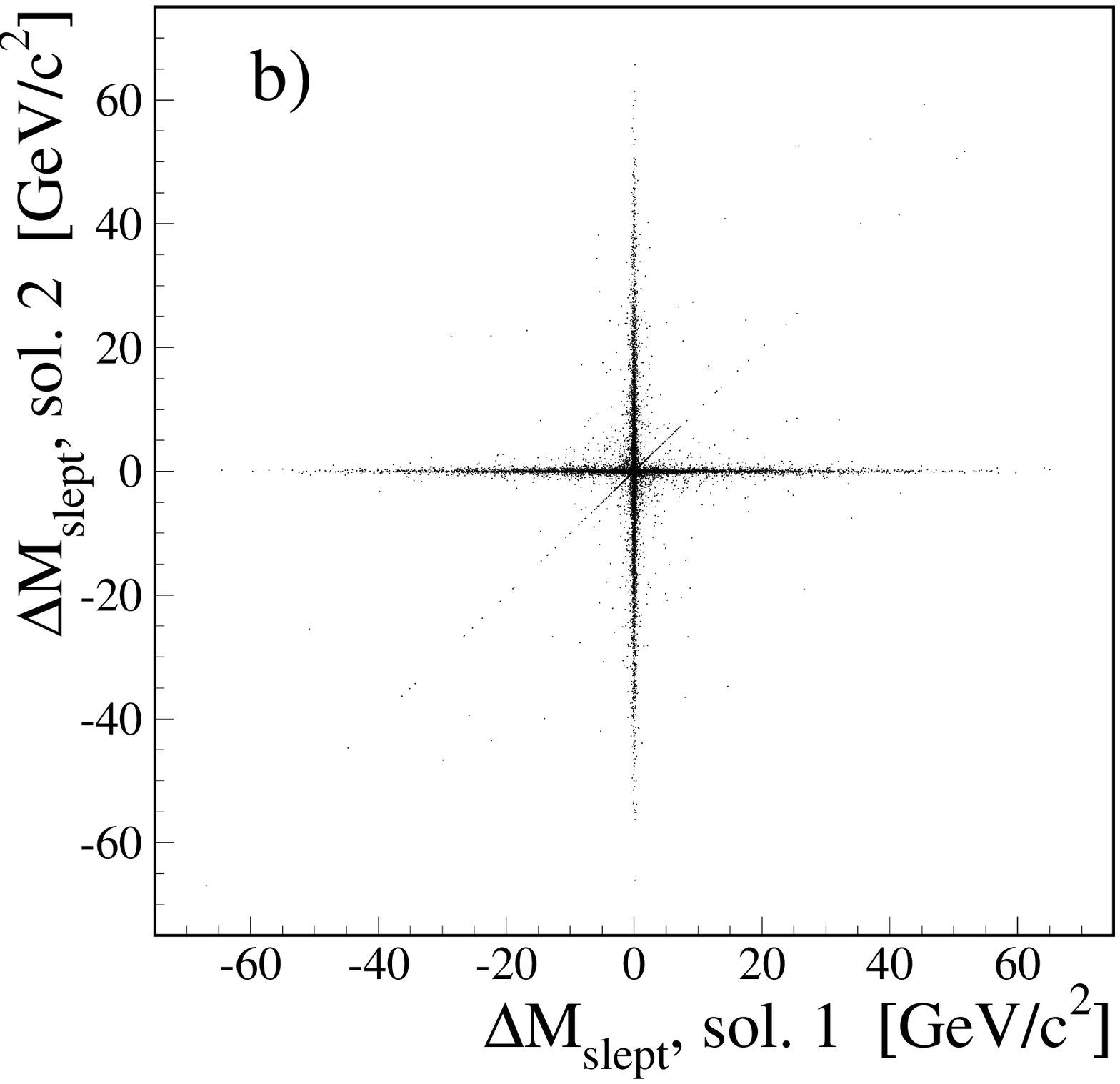}\includegraphics[%
  scale=0.18]{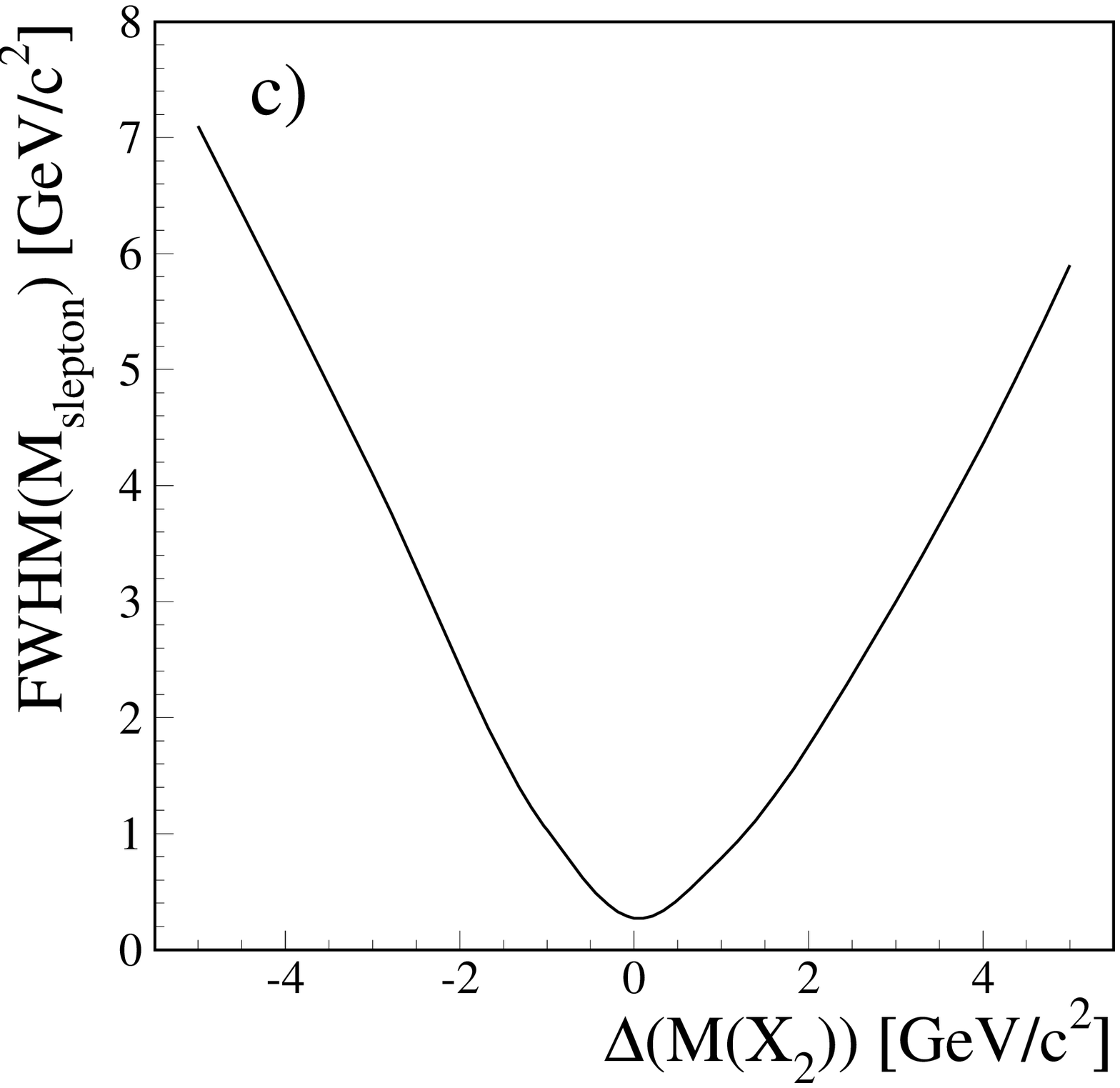}

\includegraphics[%
  scale=0.18]{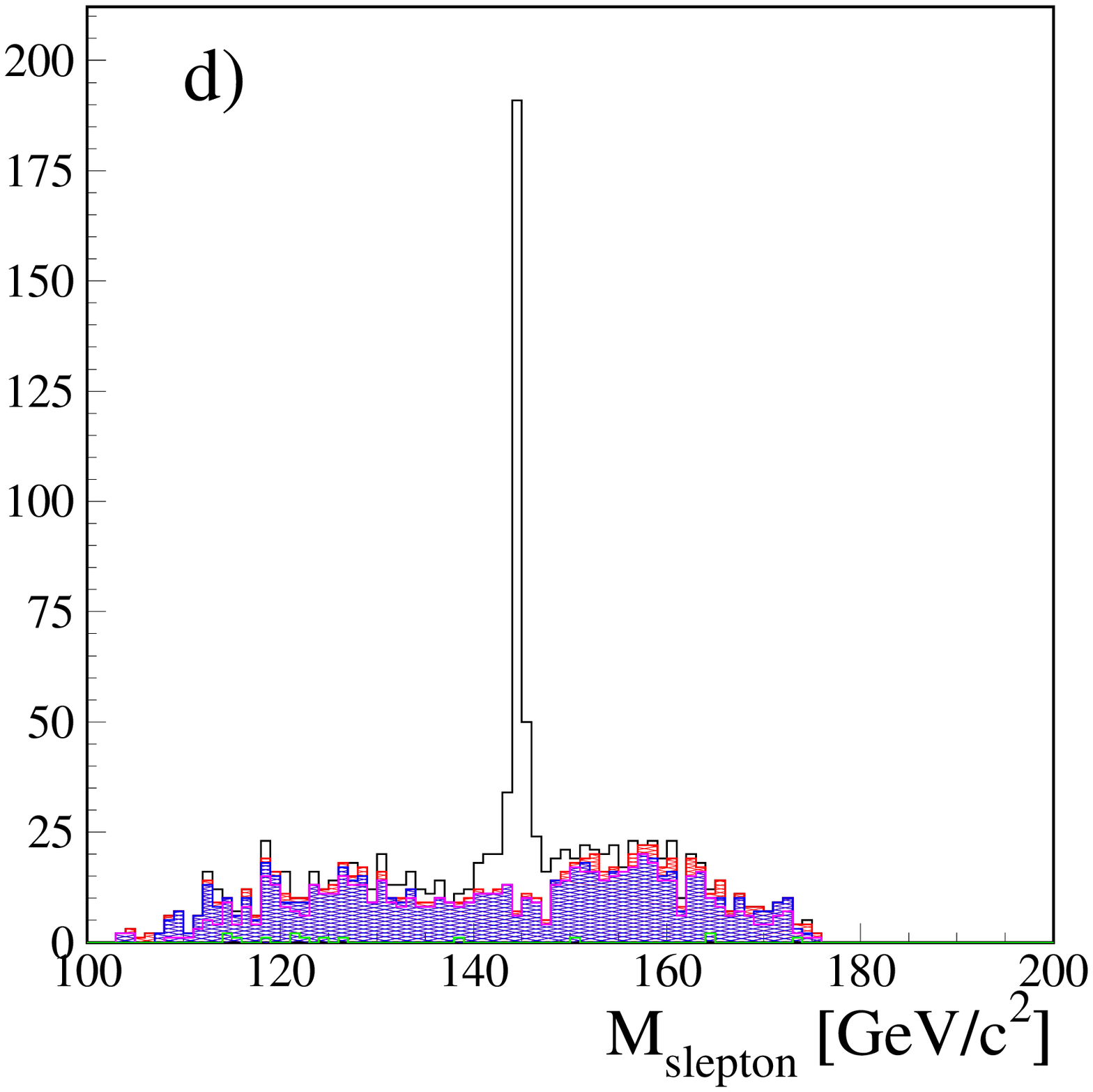}\includegraphics[%
  scale=0.18]{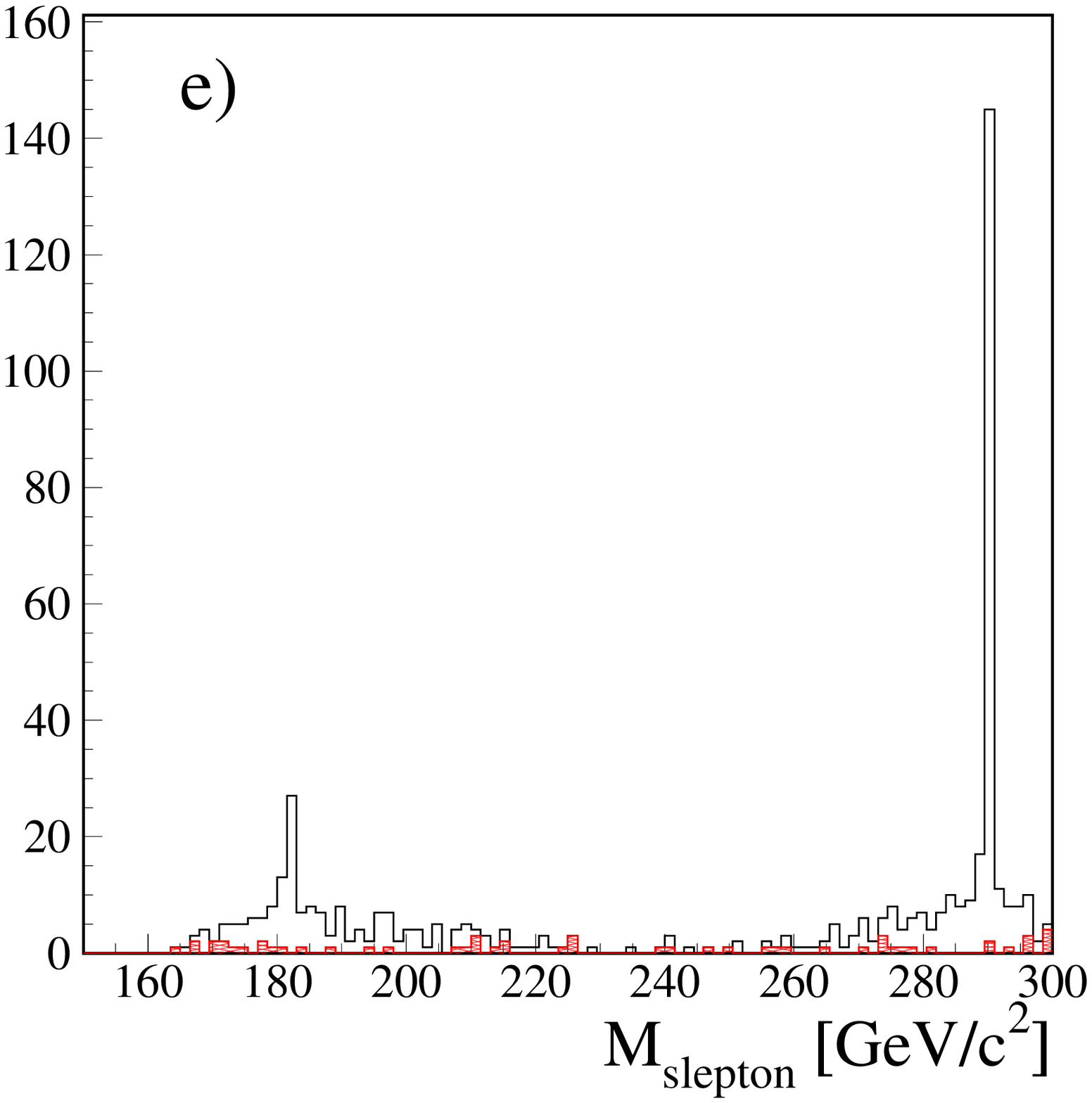}\includegraphics[%
  scale=0.18]{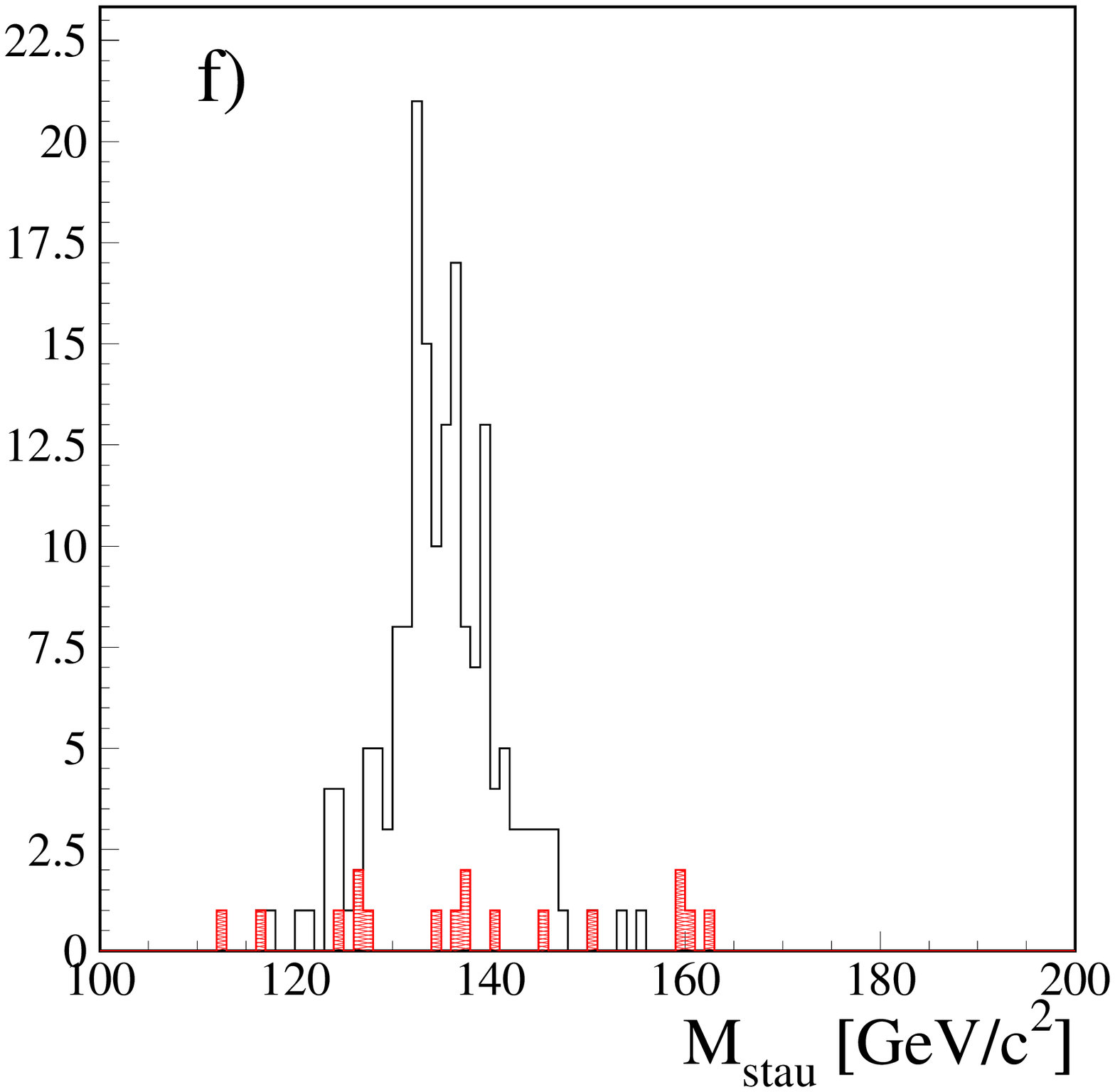}

\caption{Various kinematic distributions, for $\smu$ in SPS1a, if nothing
else is mentioned. a) Dalitz plot. b) Difference between the two reconstructed
$\msmu$ for each of the solutions. c) FWHM of the peak as a function
of $\MXN{2true}-\MXN{2input}$. d) $\msmu$. e) $\msmu$ in SPS3.
f) $\mstau$. }
\end{figure}
One can often reject the wrong \XN{2}-solution by demanding that
the masses are equal: in the vertical band in fig 1b, solution 1 is
likely to be correct, in the horizontal, solution 2. In fig 1c, the
width of the peak is shown as a function of \MXN{2}: the clear minimum
shows that $\MXN{2}$ can also be extracted from the data. Fig 1d
shows the observed distribution of the reconstructed $\msle$ with
background, ISR and beam-strahlung included (the two latter do not
effect $\sigma$, but migrates events to the tails, reducing $\epsilon$).
The background, dominated by $\selr\sell\to e\XN{1}e\XN{2}\to e\XN{1}e\sle l\to e\XN{1}ell'\XN{1}$,
is flat. The width of the reconstructed mass peak is 83 \MeVcc (experimental
resolution: the generated natural width is 0). There are 90 events
in the peak ($\epsilon$= 11 \%). The error on the mass (=$\frac{\sigma}{\sqrt{N}}$)
is 8.7 \MeVcc (0.05 \textperthousand). The fitted mass (true mass)
is 174.74 (174.73) \GeVcc. 

{\bf Multiple cascades.} In SPS3, both \selr\
and \sell\ have masses below \MXN{2}, yielding cascades that cannot
be separated by the identity of the leptons. Still, with minor modifications,
the same method can be used to extract both $\msle$. The Dalitz-triangle
does not depend on $\msle$, but the presence of several particles
will show up as more bands. To analyse the events, the position of
one peak is determined, then events with masses close to this value
in one Dalitz-projection are cut out, which enables to find the smaller
peak in the \emph{other} projection (fig 1e). There is little SUSY
background \emph{}in SPS3\emph{,} since both \selr\ and \selr\ have
masses below \MXN{2}: The $\selr\sell$ channel cannot be confused
with the signal. The fit to the two peaks yields $\sigma_{L(R)}=103(176)\MeVcc$.
At \Ecms = 800 \GeV\ and with \Lum=500 pb$^{-1}$, 50(18) events
are found in the peak, yielding $\delta(\msmu)=$14.5 (41.4) \MeVcc
(0.05 (0.2)\textperthousand). The fitted mass is 290.0(181.9) \GeVcc,
while the true value is 289.96 (181.83) \GeVcc. 

{\bf Stau channels.} Because of the $\nu$:s in the $\tau$-decays,
the method will not work for \stau:s. However, in channels where
only one \XN{2} decays to a \stau, an approximation can be found,
if \msmu (or \msel) is known: using the known masses (including $M_{\tau}$),
the the events can be reconstructed, except for two unknowns. One
- the azimuthal angle between the $\tau$ and the $\tau$-jet - has
a minor effect on the reconstructed \mstau. The other is $E_{\tau}$
in the \XN{2} decay, which can be 
estimated using all available information, and
an initial guess on \mstau. At certain combinations of $E_{jet}$
and $M_{jet}$, $E_{\tau}$ can only vary within a quite limited range,
and - under very general assumptions - the p.d.f. of $E_{\tau}$ can
be determined. Therefore, for any $E_{jet}$ and $M_{jet}$, E($E_{\tau}$)
and V($E_{\tau}$) can be found, and one can select events with low
variance estimates . The cuts used are the same as for \sel\ and
%\smu\ except that the topology should be two leptons and two $\tau$-jets\cite{delphi},
\smu\ except that the topology should be two leptons and two $\tau$-jets[10],
and that the $\selr\sell$ background must be removed by selecting
events without electrons. In fig 1f, the final mass plot for SPS1a
is shown. One finds $\sigma$= 5.07 \GeVcc, and 176 events in the
peak ($\epsilon$= 7 \%). This leads to $\delta(\mstau)=$ 380 \MeVcc
%(3 \textperthousand), comparable to other methods\cite{staus}. The
(3 \textperthousand), comparable to other methods[11]. The
fitted mass (true mass) is 135.3 (135.4) \GeVcc. $\mstau$ is one
input to the $E_{\tau}$ estimation; it was verified that the fitted
value was not sensitive to the input. By varying the $\stau$ mixing,
it was also verified that the peak indeed moves with $\mstau$. 

{\bf Conclusions.} By reconstructing $\XN{2}$ cascades
when the cascade passes a $\sle$, I show that $\delta(\msle)$ as
low as 8.7 $\MeVcc$ ($\smu$ in SPS1a) can be obtained, with ISR,
beam-strahlung, detector-resolution, background and ambiguities taken
into account. An approximate method to treat $\stau$-channels was
presented.

\end{document}